\documentclass[aps,aip,twocolumn,amsmath,amssymb]{revtex4}
\usepackage{bm}
\usepackage{amssymb}
\usepackage{amsmath}
\usepackage{latexsym}
\usepackage{graphicx}
\begin{document}
\author{Stanislav Stoupin}
\affiliation{Advanced Photon Source, Argonne National Laboratory, Illinois, USA}
\author{J. Katsoudas}
\affiliation{Illinois Institute of Technology, Chicago, Illinois, USA}
\author{Bing Shi}
\affiliation{Advanced Photon Source, Argonne National Laboratory, Illinois, USA}

\title{Flux monitoring hard X-ray optics in a single electrode configuration}

\begin{abstract}
To explore possibilities for X-ray flux monitoring on optical elements 
electrical responses of silicon and diamond single crystals and that of an X-ray mirror 
were studied under exposure to hard X-rays in a single electrode configuration in ambient air. 
To introduce flux monitoring as a non-invasive capability a platinum electrode was deposited on a small unexposed portion of the entrance surface of the crystals while for the X-ray mirror the entire mirror surface served as an electrode. 
It was found that the electrical responses are affected by photoemission and photoionization of the surrounding air. The influence of these factors was quantified using estimations of total electron yield and the ionization current. 
It is shown that both phenomena can be used for the non-invasive monitoring of hard X-ray flux on the optical elements. Relevant limits of applicability such as detection sensitivity and charge collection efficiency are identified and discussed.

\end{abstract}

\maketitle


\section{Introduction}

In hard X-ray optics monitoring flux of X-rays incident on an optical element typically requires a stand alone X-ray detector placed upstream of the optical element. Such an X-ray detector may alter the incident X-ray beam by absorbing a fraction of incident radiation and/or disturbing the radiation wavefront. Monitoring X-ray flux by the optical element can be considered as a solution to these problems. Such monitoring would particularly benefit X-ray instruments at synchrotron and X-ray free-electron laser sources where access to certain optical components is limited during an operation cycle. 

A very basic example of flux and beam position monitoring X-ray optics is a transmission window which can be operated in either a photoemission mode (electrodes on one surface) or a photoconductive mode (electrodes on the opposite surfaces)~\cite{Shu_JSR98}. 
Availability of wavefront preserving front-end X-ray windows is limited due to demanding requirements on material properties (perfect single crystal with low X-ray absorption such as beryllium or diamond). To mitigate this problem many beamlines are operated in a windowless mode (i.e., under high vacuum environment of a source). Due to the high vacuum and harsh radiation environment it can be problematic to implement stand-alone X-ray monitors for individual front-end optical components (e.g., diffracting crystals, X-ray mirrors, refractive lenses, etc.). 
Besides front-end components most hard X-ray beamlines include other optics operated in ambient air or helium environment such as focusing and collimating mirrors, capillaries, high-resolution monochromators and analyzers, zone plates and others. 
All these applications call for integration of X-ray flux monitoring capabilities into an optical element.

Such integration can be accomplished using collection of electric carriers generated in the bulk of the optical element 
(i.e., solid state detector) and/or on the surface of the optical element (photoeffect).
The first approach seems to be ideally suitable for optical elements made of semiconductor materials (e.g., 
Si, Ge, C (diamond)). However, achieving performance parameters of dedicated solid state radiation detectors 
requires tailoring of the bulk semiconductor properties (e.g., forming charge depletion regions via doping) and geometry optimization (e.g., thickness reduction). Such modification of a semiconductor material is generally not compatible with the performance characteristics of the optical element.   
For example, detection of hard X-rays using a voltage induced across a 300-$\mu$m-thick diffracting Si crystal was demonstrated earlier \cite{Afanas'ev78}. To improve X-ray detection sensitivity the crystal volume interacting with the incident X-rays was modified to form a p-n junction. However, such approach does not provide the optimal solution, since the primary requirement on the diffracting X-ray crystal optics is the optimization of crystal quality. Achieving the optimal crystal quality requires either particular dopants with specific concentrations (generally incompatible with creating p-n junctions) or the lowest possible concentration of impurities.


Bulk material properties and geometry are not critical for detector performance in the photoemission mode.   
Photoemission is one of the basic outcomes of interaction of X-rays with matter where an absorbed X-ray photon creates multiple photoionization events while some of the generated electrons leave the exposed material.
Observations of X-ray induced photoemission from diffracting crystals have been reported in literature \cite{Shchemelev73,Kikuta75,Kikuta77,Kikuta78,Afanas'ev92}. These prior studies were focused on variation of electron yield in vicinity of a Bragg diffraction condition. Such variation is due to dynamical effects in Bragg diffraction and can be reliably observed in cases when the incident radiation bandwidth and its angular spread are less or comparable to the intrinsic energy bandwidth and angular acceptance of a studied Bragg reflection. 

In the soft X-ray regime (photon energies $<$~5 keV) detection of X-rays is readily performed using electric current due to the photoemission effect. A sample or an optical element (usually conductive) is in direct contact with a conductive sample holder that is connected to the electrical ground through a current meter as shown in Fig.~\ref{fig:TEY0}.  As an uncompensated charge develops due to escape of photoelectrons a compensating electric current flows to the sample holder and is registered by the current meter.  The magnitude of this current can serve as a measure of the incident or absorbed photon flux. This measurement mode is often referred to as total electron yield since all electrons that emerge from the surface as a result of photoemission are detected, independent of their energy (e.g., \cite{deGroot_book,Ebel04,Vlachos04}). 

In this work a similar strategy was explored for realization of hard X-ray flux monitoring optical elements in a single electrode configuration. Platinum electrodes were deposited on small portions of the entrance surface of X-ray optical grade diamond and silicon single crystals (i.e., low impurity concentration and high crystal quality). The electrodes were not directly exposed to the incident radiation. Contrary to a common approach of forming a metallic electrode across the entire working surface, this configuration provides non-invasive monitoring of the incident X-ray flux since the resulting wavefront distortion are only due to the function of the optical element. It is expected that charge collection is facilitated by a drift of generated electric carriers in the lateral direction towards the electrode with an applied electric potential. 
In addition, X-ray monitoring properties of a Pd mirror (Pd film deposited on Si substrate) were explored in a similar configuration where the entire working mirror surface served as an electrode.

Although the experiments were performed in ambient air, it was found that the approach can be used to monitor X-ray flux incident on the optical elements. The results suggest that under the experimental conditions factors other than electron yield can play the dominant role.
The position sensitivity was limited due to generation of charge carriers in the surrounding air. An improvement in the position sensitivity is expected under high vacuum conditions. It was also shown that in certain cases monitoring radiation exiting an optical element becomes possible.

This paper is organized as follows. First, estimates of hard X-ray induced total electron yield are presented for basic materials used in X-ray optics and implications to X-ray flux monitoring are discussed. The obtained estimates for total electron yield illustrate the limits of hard X-ray detection sensitivity in the photoemission mode. In the next step, several experiments using synchrotron radiation are described where X-ray monitoring in the single electrode configuration was demonstrated for X-ray optical grade diffracting crystals and an X-ray mirror in ambient air.
The estimates for total electron yield are used to benchmark the experimental data collected under different experimental conditions. In particular, it is shown that photoionization of a surrounding medium can be used for monitoring X-ray flux on the X-ray mirror operated in an enclosed environment under moderate-to-low incident hard X-ray flux, while the contribution of the total electron yield can be neglected.
Finally, these and other experimental observations are discussed and summarized.

\section{Total Electron Yield}\label{sec:TEY}


\begin{figure}
\setlength{\unitlength}{\textwidth}
\centering\includegraphics[width=0.5\textwidth]{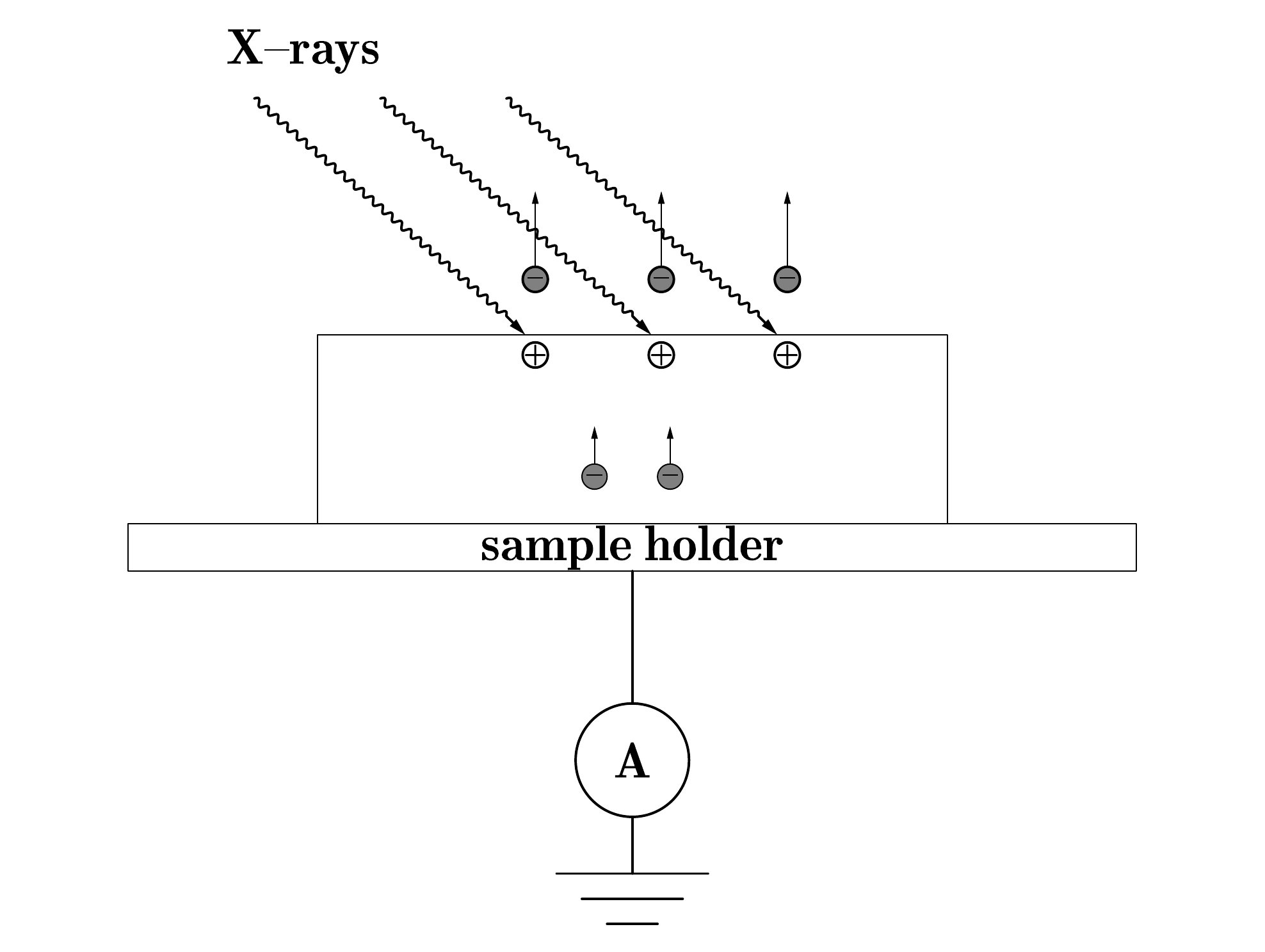}
\caption{Basic experimental configuration for measurement of total electron yield by a compensating electric current flowing to the sample in soft X-ray spectroscopy.}
\label{fig:TEY0}
\end{figure} 


In the total electron yield configuration (Fig.~\ref{fig:TEY0}), bulk conductivity in the sample promotes compensation of the electric charge, thus improving detection sensitivity. 
If the electric conductivity is limited charge compensation may become incomplete and a potential barrier develops that prevents further escape of the electrons. Typical solutions to this problem are modification of bulk electrical conductivity of the sample (e.g., mixing with electrically conductive material) and/or reducing the electric potential of the sample using a voltage source. An additional electrode (e.g., a grid with a positive potential located above the surface of the optical element) can be used to accelerate the escaping electrons and prevent their return to the surface. 

\begin{figure}
\setlength{\unitlength}{\textwidth}
\centering\includegraphics[width=0.5\textwidth]{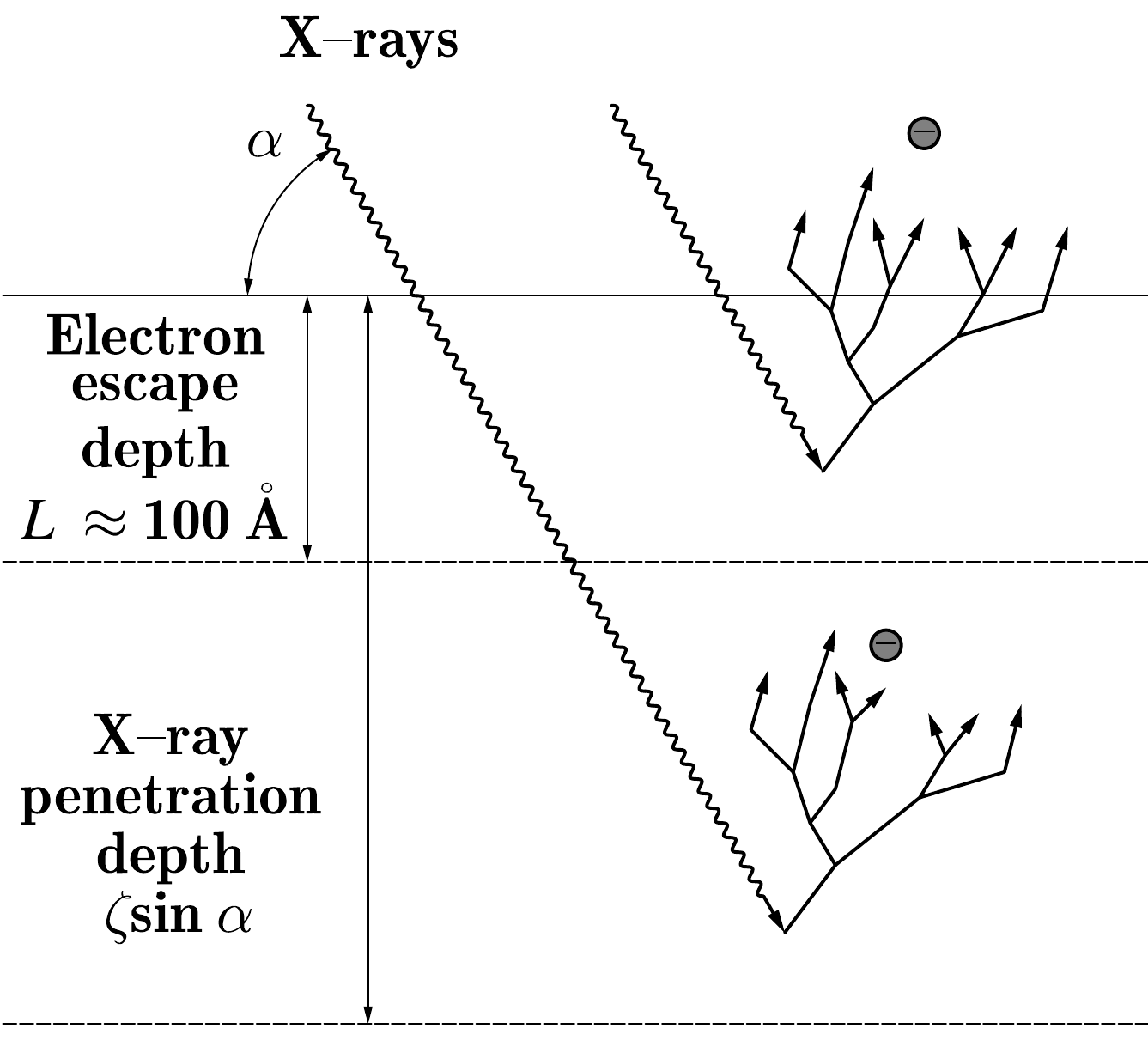}
\caption{Photoabsorption and electron production at photon penetration depth (electrons do not escape the material) and within the electron escape depth (IMFP) (electrons can escape the material).}
\label{fig:photoab}
\end{figure}

Incident X-ray photons penetrate into the bulk of the material to a characteristic depth $\zeta \sin{\alpha}$, 
where $\zeta$ is the X-ray absorption length given by the inverse of the linear attenuation coefficient $\mu$ [cm$^-1$], which is a function of the photon energy $E_X$; and $\alpha$ is the glancing angle of incidence to the sample surface as shown in Fig.~\ref{fig:photoab}.

The dominant contribution to attenuation of X-rays (in the most practical range for X-ray optics $E_X \lesssim$~30~keV)
is due to the photoelectric absorption cross section $\sigma^{pe}$, 
which is a measure of probability of photoionization (i.e., creation of a photoelectron upon absorption of an X-ray photon). Thus, in our practical consideration $\mu \simeq \rho_n \sigma^{pe} $, where where $\rho_n$ is the atomic volume density [atoms/cm$^3$].
\footnote{For elements with low atomic number such as Be or C scattering cross sections become comparable to $\sigma^{pe}$
 at photon energies $E_X \approx$~15~keV.} 

A fraction of excited photoelectrons may completely escape the material. On the way to the surface these electrons exhibit inelastic scattering events, which results in the reduction of the energy of the primary photoelectrons. Electrons generated deeply in the bulk of the material have insufficient energy to escape. The electrons can escape the material only if they are generated within a certain characteristic depth known as electron inelastic mean free path (IMFP), which is a function of the photoelectron energy.  For many elementary materials IMFP values have been calculated, experimentally verified and compiled into a database \cite{NIST71} (a relatively small number of experimental studies have been performed at hard X-ray photon energies).
Inelastic scattering does not only reduce the energy of the primary photoelectrons but also produces a cascade of secondary electrons with smaller energies as shown schematically in Fig.~\ref{fig:photoab}. 
Secondary electrons comprise a major (dominant) portion of the total electron yield \cite{Eliseenko68,Henke81}.
In addition, Auger electrons of certain characteristic energies are created as a possible de-excitation route for the absorbing atom. 
Auger electrons can also escape the material thus contributing to the total electron yield. 

A simplified analytical description of such complicated process can be offered based on several assumptions arising from experimental observations as derived in \cite{Stohr_book}. An assumption is made that the energy distribution of low-energy secondary electrons is independent on the primary electron energy once it is higher than about 20~eV and that the number of the secondary electrons is proportional to the incident photon energy $E_X$. The electron gain factor (number of electrons generated per one photoionization event) is $G^e=E_X M$, where $M$ is a material constant describing the conversion efficiency. In analogy to attenuation of X-rays a quantity $1/L$ is introduced as a linear electron-attenuation coefficient (where $L$~$\approx$~100~$\rm \AA$ is the effective energy-independent electron escape depth) that describes the electron scattering process as an attenuation of a single primary photoelectron multiplied by the gain factor $G^{e}$.  

For a semi-infinite slab of material, the number of photoelectrons emitted per single incident X-ray photon of energy $E_X$ (total quantum yield) is given by 
\begin{equation}
Q = \frac{1}{2} (1-R(\alpha)) G^e \frac{L}{\zeta \sin{\alpha} + L},
\label{eq:y1}
\end{equation} 
where $R(\alpha)$ is the material reflectivity for the incident radiation.

In grazing incidence under the condition of total external reflection (i.e., X-ray mirror case) a substantial increase in the 
quantum yield is expected. This condition is satisfied if $\alpha < \alpha_C$, where $\alpha_C$ is the critical angle that depends on the choice of the material and incident photon energy (e.g., \cite{ANMM_book}).
An estimate of total electron yield can be performed by replacing in Eq.~\ref{eq:y1} the x-ray penetration depth $\zeta \sin{\alpha}$ with an X-ray attenuation length $\Lambda$ in total external reflection. The values of $\Lambda$ (see e.g., an online calculator \cite{CXRO}) can be several times smaller than the effective electron escape length (i.e., $\Lambda \ll L$).  Under this approximation the quantum yield does not depend on the photoelectric absorption in the material:
\begin{equation}
Q \simeq \frac{1}{2} (1-R(\alpha)) G^e .
\label{eq:qy1}
\end{equation}

Although, only a small fraction of the incident X-ray flux $(1-R(\alpha)) \simeq 10^{-2}$ can contribute to photoelectric absorption in total external reflection, the smallness of the penetration depth provides an enhancement. This is due to the fact that in total external reflection X-rays propagate nearly parallel to the surface of the material and interact mostly with electrons contained within the effective photoelectron escape depth. 
The primary photoelectrons, Auger electrons and the secondary electrons produced near the surface have higher probability to escape the material which leads to enhancement of total electron yield.

In absence of total external reflection the absorption depth is much larger than the effective electron escape depth
$\zeta \sin{\alpha} \gg L$. The quantum yield is given by
\begin{equation}
Q \simeq \frac{1}{2} (1-R(\alpha)) G^e \frac{L}{\zeta\sin{\alpha}}.
\label{eq:qy2}
\end{equation} 

It is thus reduced by the ratio of the effective photoelectron escape depth to the effective photoabsorption penetration depth. 
This illustrates a non-invasive character of X-ray flux monitoring in the photoemission mode at the expense of reduction in detection sensitivity.
Many cases that fall within this scenario do not exhibit substantial reflection (i.e., $(1-R(\alpha)) \simeq$~1). These include refractive optics and primary diffracting optics where only a small fraction of incident radiation with a wide energy/angular distribution can be reflected into a narrow energy/angular range of the diffracting element.

Table~\ref{tab:param} shows a summary on quantum yield estimated using Eq.~(\ref{eq:y1}) 
for several X-ray optical materials using experimental data available in the literature.
In case of Au the value for $G^e$ was derived in \cite{Henke81}. It was shown that Eq.~\ref{eq:y1} closely approximates the measured total electron yield for Au even in vicinity of 10~keV. For C and Si the values for electron yield relative to that of Au were given in the literature for particular energies. In these cases $G^e$ was calculated at those energies and extrapolated to a representative photon energy of 10~keV assuming the linear dependence of $G^e$  on the photon energy.

In the case of total external reflection (TER) for Au the quantum yield can be as high as Q~$\simeq$~0.6, which is due to the fact that the penetration depth is only about 10 $\rm{\AA}$ (much less than $L = 50 \rm{\AA}$ for Au~\cite{Stohr_book}). 

In absence of total external reflection higher quantum yields can be obtained if more photons are absorbed within the effective photoelectron escape depth (i.e., for materials with greater photoabsorption) as follows from Eq.~\ref{eq:qy2}. To obtain representative numbers in these cases normal incidence was assumed, i.e. $\alpha = \pi/2$. 


The dominant factor that affects the quantum yield is the photoabsorption length (assuming a relatively small variation in the effective electron escape depth $L$). Therefore, the key cases given in Table~\ref{tab:param} approximate other materials commonly used in hard X-ray optics. Silicon and diamond single crystals represent many of cases in diffractive optics. Quantum yield for other diffracting crystals such as Ge, Al$_2$O$_3$ (sapphire), SiC and SiO$_2$ (quartz) should be on the same order of magnitude as that of Si. 
The main material for refractive optics is Be. Here, due to reduction in photoabsorption (low-Z material) the quantum yield should be smaller than that of diamond. However, there are many practical examples of refractive lenses made out of Si and diamond. Finally, it is expected that for thin films of Pt, Pd, Rh, etc. used as X-ray mirrors the TER quantum yield is similar to that of Au (no strong dependence on the photoionization cross-section and only small variation in the penetration depth). 

\begin{table}
\caption{Electron gain factor and quantum yield for representative materials used in hard X-ray optics: in absence of total external reflection with $\alpha = \pi/2$ and in the case of total external reflection at $\alpha \simeq$~2.5~mrad (TER). The experimental values for the electron gain factor $G^e(E_0)$ at photon energy $E_0$ are extrapolated to $E_X$~=~10~keV (using linear dependence of $G^e$ on the photon energy). Estimates of maximum possible electric current $I$ due to total electron yield with incident photon flux $F$~=~$1 \times 10^{16}$ photons/s are given in the last column.}
\begin{tabular}{l c c c c c c c}
\hline\hline
material  & $G^e(E_0)$ &$E_0$   &Ref.              &$G^e(E)$      &$Q(\alpha)$     &$I(10^{16})$ \\
          &            &(keV)   &                  &$E$=10~keV     &$E$=10~keV                &$\mu$A \\
\hline
diamond   &4.3         &1      &\cite{Kummer13}    &43         &1.4e-4             & 0.23  \\         
Si        &3.8         &0.16   &\cite{Kitamura95}  &236        &1.0e-2               & 16 \\         
Au (TER)  &4.3         &1.5    &\cite{Henke81}     &29         &0.6                & 912 \\  
\hline\hline
\end{tabular}
\label{tab:param}
\end{table}

The last column in the table represent an electric current that would completely compensate the electric charge due to electron escape events: 
\begin{equation}
I = q^e F Q ,
\label{eq:i}
\end{equation}
where $q^e \simeq 1.6 \times 10^{-19}$~C is the electron charge and $F$ is the photon flux.
The values given in the column correspond to the photon flux $F \simeq 1 \times 10^{16}$ photons/s delivered by an undulator at a third-generation synchrotron through an aperture of about 1~$\times$~1~mm$^{2}$ at a typical source-to-aperture distance of $\approx$30~m. 
For comparison, a photon flux generated by a laboratory source of X-rays (e.g., an X-ray tube) $F \approx 1 \times 10^{10}$ photons/s emitted in a solid angle of a few millisteradians. Thus, for the laboratory source the estimated electric currents are six orders of magnitude less (i.e., the same values in the units of pA).
These two cases can be considered as a full dynamic range for monitoring capabilities of the X-ray optical element using total electron yield. Detection of small electric currents is feasible down to sub-pA regime (although below a few pA specialized procedures to reduce noise are required). 
Thus, monitoring incident flux on X-ray optical element using total electron yield is feasible even in systems that utilize conventional X-ray sources such as those in medical or analytical X-ray laboratory. The other end of the dynamic range with expected currents in the $\mu$A regime represent front-end optical components of a synchrotron beamline. In the following these estimates of total electron yield will be used for benchmarking the experimental data.

%

\section{Experiments, Results, and Interpretation}
\begin{figure}
\setlength{\unitlength}{\textwidth}
\centering\includegraphics[width=0.5\textwidth]{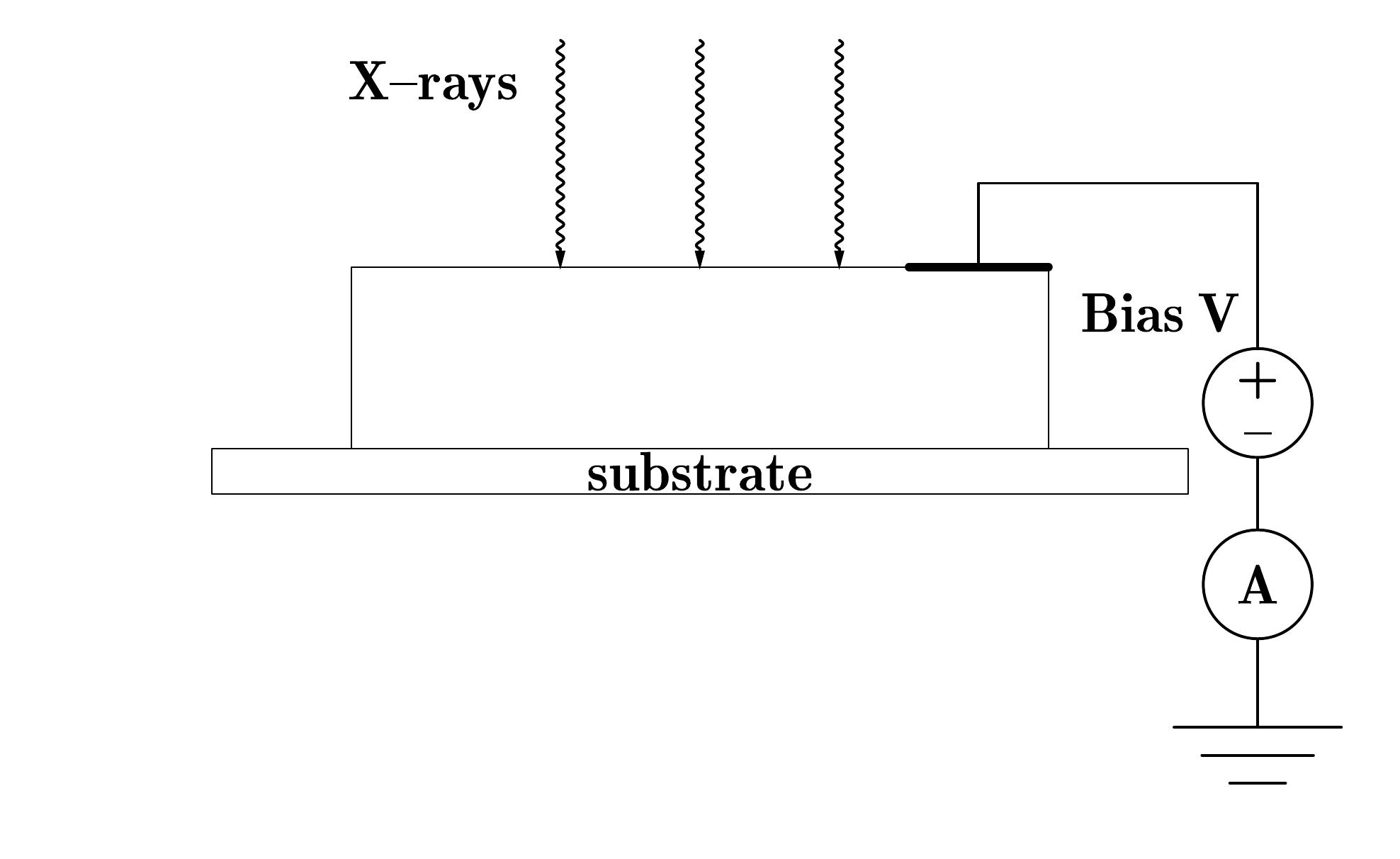}
\caption{Experimental configuration for electric measurements on X-ray optical quality single crystals. A single Pt electrode is deposited on the optical surface of the crystal. The electrode is grounded through a current meter. A bias voltage is applied to the electrode with respect to the ground.}
\label{fig:TEY1}
\end{figure}

Platinum electrodes were deposited on small sections of working surface of type IIa diamond crystal plate and a high resistivity Si crystal. The dimensions of the diamond plate were approximately 5$\times$6$\times$0.3~mm$^3$ and the dimensions of the Si crystal were 20$\times$40$\times$10~mm$^3$. The crystals were placed on a non-conductive substrate. For a diamond crystal the substrate was a 0.5-mm-thick CVD diamond plate having a rectangular window of 5$\times$2~mm$^2$ (a construction similar to the all-diamond X-ray optical assembly~\cite{Stoupin14}).
For the Si crystal the substrate was a 0.5-mm-thick Kapton\textregistered{} sheet. 
The experimental configuration for measuring an X-ray induced electric current is shown in Fig.~\ref{fig:TEY1}.
The experiments were conducted in ambient air at the MRCAT 10BM (bending magnet) beamline~\cite{Kropf_AIP10} of the Advanced Photon Source (Argonne National Laboratory, Illinois). 

\begin{figure}
\setlength{\unitlength}{\textwidth}
\centering\includegraphics[width=0.5\textwidth]{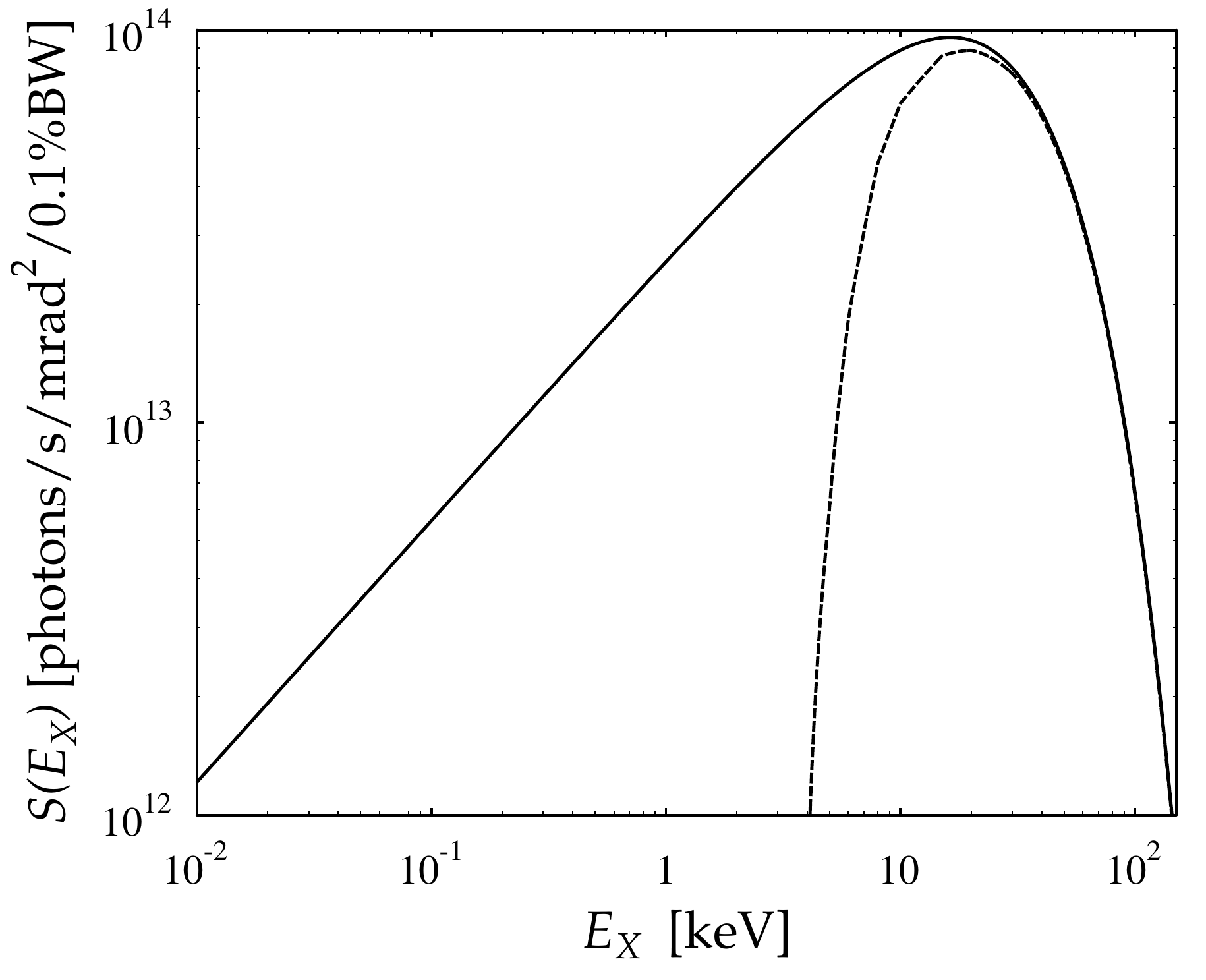}
\caption{The spectral flux of the bending magnet radiation. The solid line shows the theoretical calculation using the well-known equations for bending magnet radiation~\cite{xdb2001}. The dashed line shows the same spectral flux attenuated by 0.5~mm-thick beryllium and 40~cm of air at the position of the optical element. }
\label{fig:sflux}
\end{figure}

In the first experiment a white (polychromatic) X-ray beam was used to maximize X-ray flux incident on the optical elements. The spectral flux $S(E_X)$ of the bending magnet radiation is shown in Fig.~\ref{fig:sflux}.
The solid line shows the theoretical calculation using the well-known equations for bending magnet radiation~\cite{xdb2001}. The dashed line shows the same spectral flux attenuated by 0.5~mm-thick 
combined thickness of beryllium windows at the beamline and 40 cm of air at the position of the optical element.

\begin{figure}
\setlength{\unitlength}{\textwidth}
\centering\includegraphics[width=0.5\textwidth]{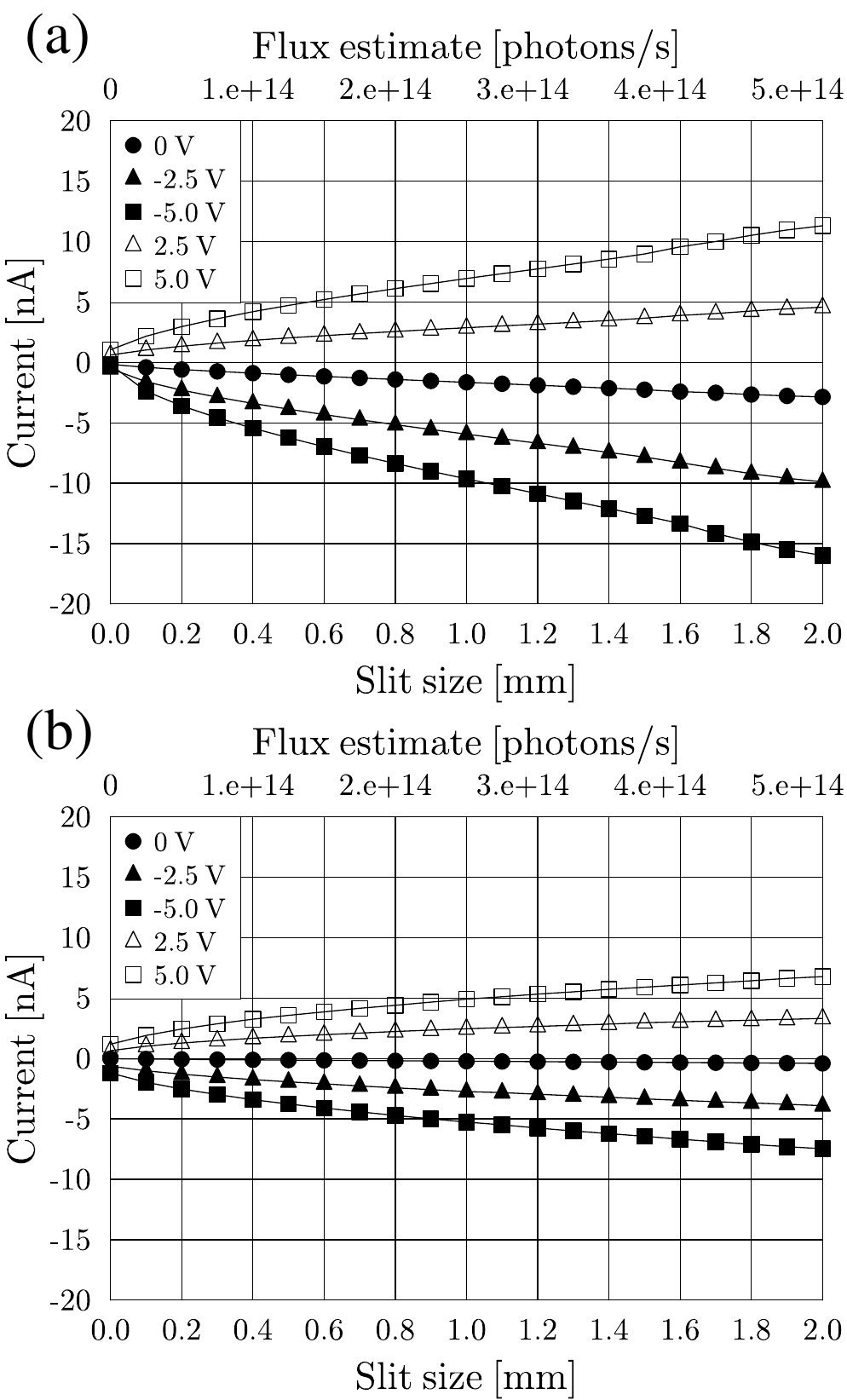}
\caption{Electric current in circuit configuration of Fig.~\ref{fig:TEY1} as a function of the horizontal size of a polychromatic X-ray beam incident on diamond crystal (a) and Si crystal (b) at different values of the applied bias voltage. The secondary x-axis shows the value of the incident X-ray flux estimated by integration of the spectral flux of the bending magnet radiation.}
\label{fig:CSi}
\end{figure}

The X-ray beam of a size up to 1$\times$2~mm$^2$ (vertical~$\times$~horizontal) was incident on the optical surface of the crystals at normal incidence ($\alpha \simeq$ 90~deg). The Pt electrodes were not directly exposed to the incident beam.
The electric current was measured as a function of the applied bias voltage and the horizontal beam size. The values of the electric current in the absence of X-rays (i.e., dark current) were subtracted from the signal. 
The observed values of the dark current were at the level of $\approx$~0.2~nA and did not change substantially with the applied voltage. The results of the measurements are summarized in Fig.~\ref{fig:CSi}. 

The variation in the beam size results in the variation of the incident flux. 
It was assumed that the X-ray illumination over the maximum aperture size (1$\times$2~mm$^2$) was uniform (i.e., the incident flux was proportional to the illuminated area or to the horizontal beam size). The flux of the incident white beam was estimated by integration of the spectral flux (Fig.~\ref{fig:sflux}). The estimated values are given by the secondary x-axis in Fig.~\ref{fig:CSi}. 
Different curves correspond to different values of the applied bias voltage (from -5~V to +5~V) as shown in the figure legend. The positive sign of the bias voltage is as show in Fig.~\ref{fig:TEY1}. 
The results show that the electric current is a monotonically increasing function of X-ray flux and bias voltage.  
Thus, if properly calibrated, this configuration can be utilized for monitoring of the incident X-ray flux. However, realization of a practical device requires detailed understanding of origins of the observed effect as well as detailed characterization of signal stability and sensitivity.
Greater values of the electric current were measured for the diamond crystal as compared the to Si crystal under the same conditions. A feasible explanation for this effect is a poor charge compensation and collection for Si where the electrode was further from the exposed area. 
Reversing bias polarity does not suppress the electric current. The suppression is expected for the pure case of electron yield where photoelectrons originating from the optical element are the only carriers. 
This observation indicates that the total electron yield is not the dominant factor that defines the electrical response of the system. At the same time, the response is notably asymmetric (enhancement at negative bias voltages) and essentially non-zero at $V=0$ (especially in the case of diamond). This observation suggests that the photoemission indeed contributes to the electric current.

\begin{figure*}
\setlength{\unitlength}{\textwidth}
\centering\includegraphics[width=0.9\textwidth]{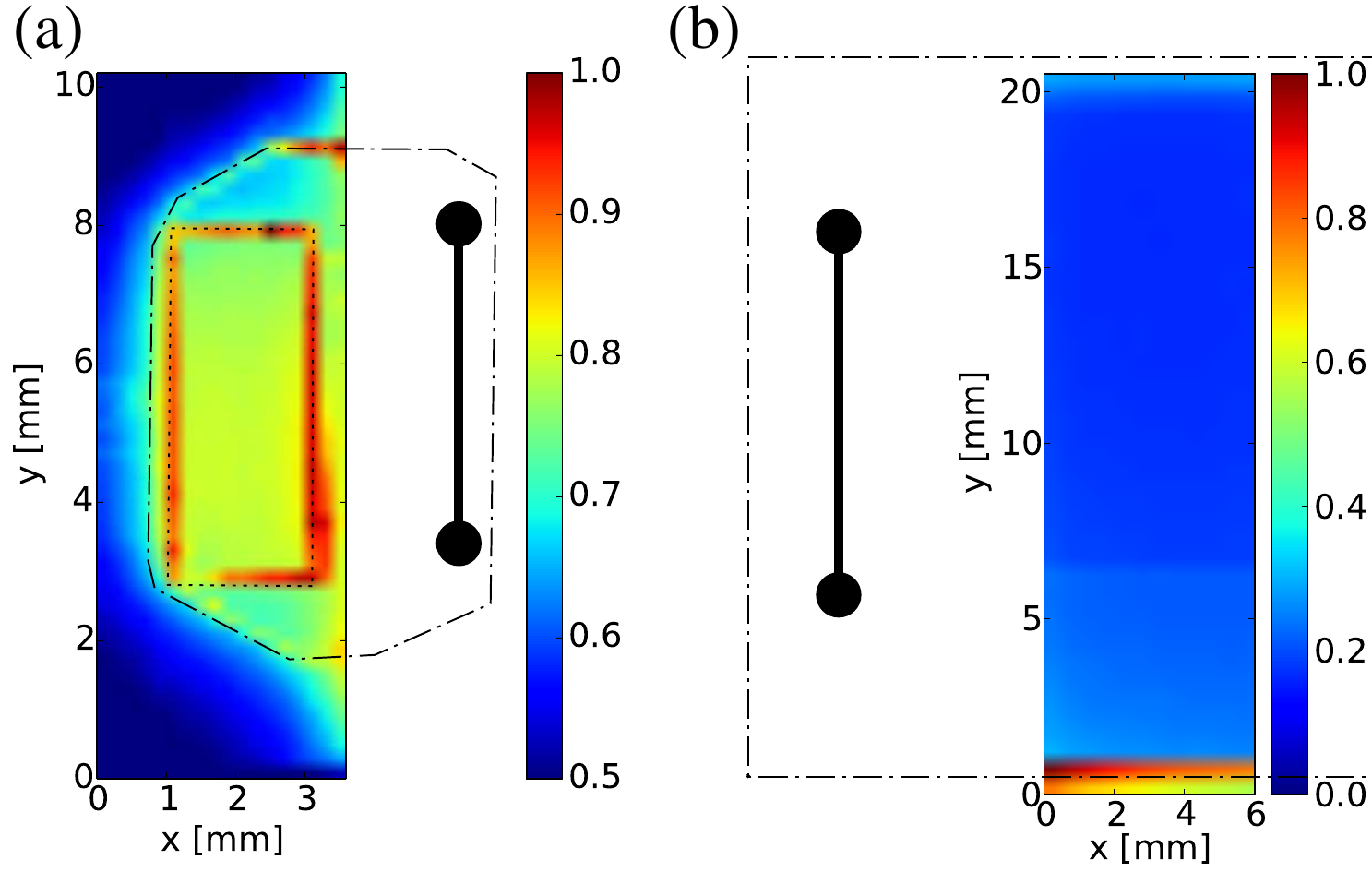}
\caption{Maps of the electric current (normalized) produced by raster scans of a section of the diamond crystal (a) and a section of the Si crystal (b). 
The outlines of the crystals are shown by dot-dashed line. The 5~$\times$~2~mm$^2$-sized window in the substrate behind the diamond crystal is shown by dashed line. The electrodes are shown by solid lines ending with circles (contact areas). 
The rasters scans were performed using an incident white (polychromatic) X-ray beam with a small cross section.}
\label{fig:map}
\end{figure*}

In the second experiment a major portion of the diamond crystal was raster scanned across a white beam with a small $0.2\times0.2$~mm$^2$ cross section size, and a section of the Si crystal was raster scanned across a white beam with $0.5\times0.5$~mm$^2$ cross section size. 
The electric current was measured during the raster scans with an applied bias voltage of -5~V. 
The resulting raster scans are shown in Fig.~\ref{fig:map}. 

These raster scans reveal that the current increases when the edges of the crystal are exposed to the X-ray beam. 
A possible explanation is that the beam is in grazing incidence to the edge surfaces, which results in an increase of electron photoemission.
The increase is particularly prominent for the Si crystal which might be related to the fact that the surface of the edge is larger due to the greater crystal thickness (Fig.~\ref{fig:map}(b)). The increase is also observed on the edges of the 5~$\times$~2~mm$^2$-sized window in the substrate behind the diamond crystal (Fig.~\ref{fig:map}(a)). 
At the same time, the contrast in the signal when the beam is on and off the crystals was not particularly strong. For example, in the case of the diamond crystal (Fig.~\ref{fig:map}(a)) the maximum current value observed at the edges was only about two times greater than the recorded current when the beam was off the crystal. A similar effect was observed for the Si crystal (not shown in Fig.~\ref{fig:map}(b)). 
This fact along with the prior observation of the greater current for diamond in comparison to Si clearly indicate that there are additional factors besides total electron yield that contribute to detection sensitivity. Additional charge carriers are generated due to ionization of air surrounding the crystals which can contribute to the electric current in the circuit. In other words, the studied configuration can act as an ionization chamber with a pair of non-well-defined electrodes. 

\begin{figure}[!h]
\setlength{\unitlength}{\textwidth}
\centering\includegraphics[width=0.5\textwidth]{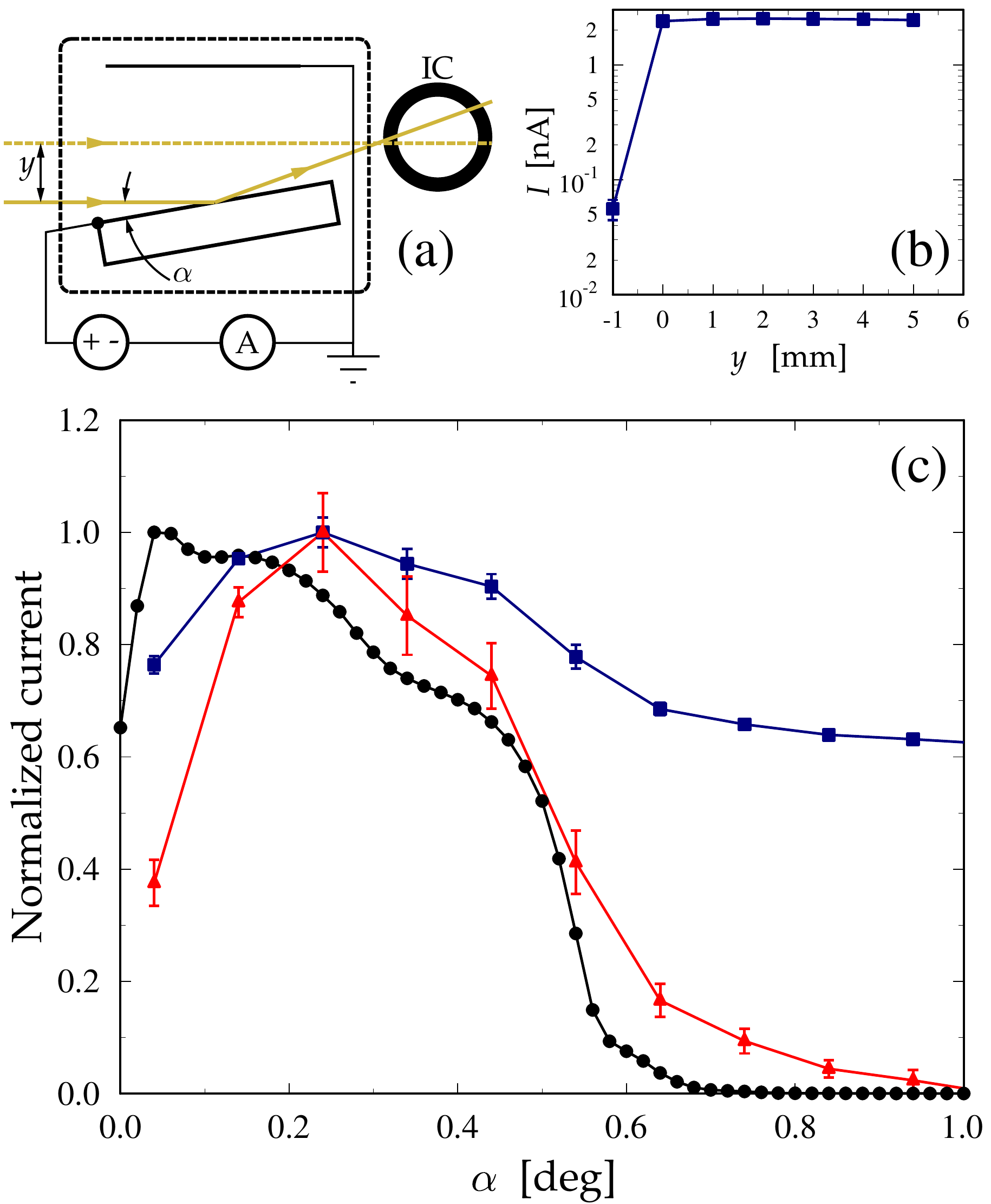}
\caption{(a) Scheme of the Pd mirror experiment (see text for details). (b) The electric current as a function of the vertical position of the X-ray beam. (c) Normalized intensity of the reflected beam measured using the ionization chamber (black line, black circles) and the normalized electric current (blue line, blue squares) as functions of the incident angle. The red line and red triangles represent the electric current normalized after subtraction of a baseline (the baseline is approximated by the current value at $\alpha \simeq$~1~deg).}
\label{fig:Pd}
\end{figure}

To explore such a possibility a third experiment was performed in which a Pd mirror (a Pd metallic film deposited on a flat Si substrate) was contained in a sealed environment. The length of the mirror was 100~mm.
The entire mirror surface served as the electrode of the optical element while a separate grounded electrode was placed above the mirror as shown in Fig.~\ref{fig:Pd}(a). The experiment was performed at an X-ray photon energy of 7.4~keV using an X-ray beam delivered by a Si (111) double-crystal monochromator. The size of the beam was $0.1\times3.0$~mm$^2$ (vertical$\times$horizontal) and the photon flux was $\approx 2 \times 10^{8}$ photons/s. The estimations presented in Sec.~\ref{sec:TEY} suggest that in such regime the electric current due to total electron yield (grazing incidence) should be on the order of few tens of pA. 

A constant bias voltage $V_B$~=~-100~V was applied initially. The intensity of the X-ray beam reflected off the mirror in grazing incidence was measured using an ionization chamber (IC) as a function of the incidence angle $\alpha$. The electric current in a circuit shown in Fig.~\ref{fig:Pd}(a) was measured as a function of the vertical position of the X-ray beam $y$ and as a function of the incidence angle~$\alpha$.

The electric current as a function of the vertical beam position is shown in Fig.~\ref{fig:Pd}(b). The position $y>0$ corresponds to X-ray beam passing above the mirror surface, $y=0$ corresponds to grazing incidence at $\alpha \simeq$~0.15~deg and $y<0$ corresponds to a situation when the incident beam is blocked by the side of the mirror substrate (i.e., the beam is below the mirror surface). A substantial increase in the electric current is observed at $y=0$ and its value remains about the same for $y>0$. This confirms that majority of charge carriers are generated in air surrounding the mirror and the contribution of photoelectrons originating at the mirror surface is small. Nevertheless, the electric current in this configuration can be used for monitoring intensity of the beam reflected from the mirror in grazing incidence. 
Figure~\ref{fig:Pd}(c) shows an angular dependence of the normalized intensity of the reflected beam measured using the ionization chamber (black line, black circles) and the normalized electric current (blue line, blue squares). The red line and red triangles represent baseline subtracted and normalized values of the electric current (the baseline is approximated by the current value at $\alpha \simeq$~1~deg). The falling edge of this curve coincides with the falling edge of the reflectivity curve at $\alpha_c \simeq$~0.5~deg (the critical angle for a thick Pd mirror at 7.4~keV). 
Such behaviour of the electric current is understood from the fact that above the critical angle of the mirror the intensity of the reflected beam falls rapidly and the number of ionization events in air created by the reflected beam decreases while those originated from the incident beam are still present.
As the incidence angle starts to approach zero the beam becomes nearly parallel to the surface. The electric current starts to decrease at $\alpha \approx$ 0.1~deg when the expected length of the beam footprint on the mirror surface begins to exceed the length of the mirror. 

\begin{figure}[!h]
\setlength{\unitlength}{\textwidth}
\centering\includegraphics[width=0.5\textwidth]{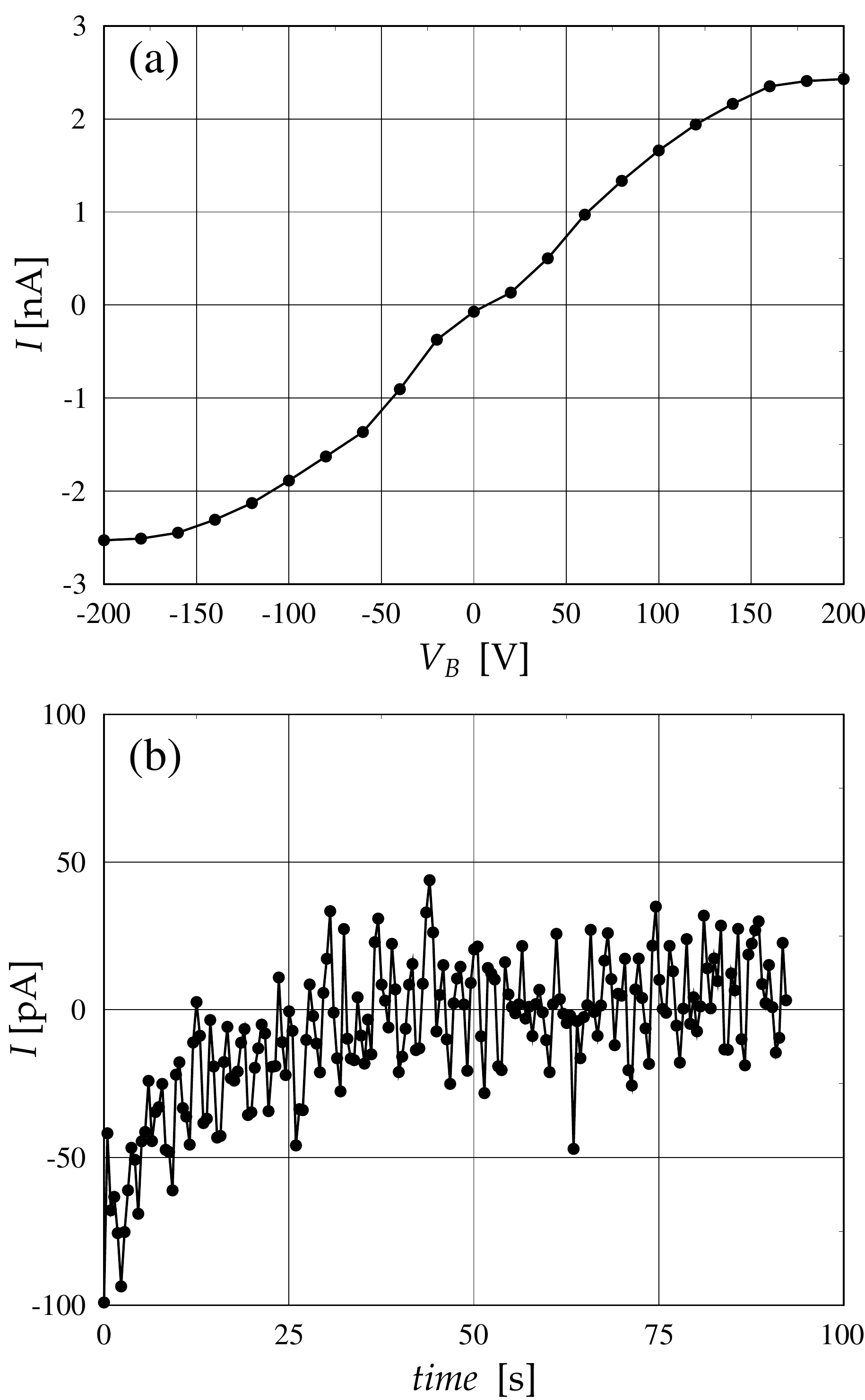}
\caption{(a) The IV curve of the enclosed X-ray mirror in grazing incidence of hard X-rays ($E_X$=7.4~keV). (b) The dark current as a function of time upon application of a bias voltage ($V_B$~=~-100~V).}
\label{fig:Pd_IV}
\end{figure}

The response of the enclosed X-ray mirror to the incident X-ray flux as a function of the the applied bias voltage (i.e., IV curve) is shown in Fig.\ref{fig:Pd_IV}(a). The response substantially exceeds the dark current. The dark current is shown in Fig. \ref{fig:Pd_IV}(b) as a function of time upon application of the -100~V bias voltage. 

The electric current in the IV curve shows signs of saturation at high levels of the bias voltage ($|V| \approx$~200~V), consistent with the behavior of an ionization chamber where an increase in the applied potential eventually leads to a substantial reduction in the recombination of the charge carriers (i.e., efficient charge collection - see e.g., \cite{Tsoulfanidis_book}). 

The saturation current can be estimated using the maximum possible number of gas ionization events produced by an X-ray photon of energy $E_X$:
\begin{equation}
I_s \simeq q^e F_a \frac{E_X}{E_i} ,
\label{eq:IC}
\end{equation}

where $F_a$ is the photon flux absorbed in the mirror chamber and $E_i$ is the gas ionization energy ($E_i \simeq$~14.5 for nitrogen). Using the parameters of the experiment Eq.~\ref{eq:IC} yields $I_s \simeq 2.7$~nA, which is close to the observed value. 

Overall, the experiment with the X-ray mirror in an enclosed configuration in the presence of an ionizable gas shows that the induced electric current is not affected by the photoemission. This is in agreement with our estimations of the total electron yield at the moderate hard X-ray flux conditions of the experiment. Instead, the enclosed configuration is naturally suited for monitoring of the X-ray flux as an ionization chamber as confirmed in the experiment.

\section{Discussion and Summary}
The conducted experiments demonstrate that a simple electric circuit like the one depicted in Fig.~\ref{fig:TEY1} can be used to monitor X-ray flux incident on an X-ray optical element. Although, the circuit with a single electrode resemble a recipe for signal acquisition in the "photoemission" mode our analysis suggests that the electric current predominantly originates from ionization of a medium surrounding the optical element. Indeed, the observed values of the electric current exceed those expected from total electron yield (Table~\ref{tab:param}). 
This, however, does not preclude the use of the circuit for monitoring the incident flux (and, in certain cases, the reflected intensity as demonstrated by the experiment with the X-ray mirror) but implies reduced position sensitivity and increased susceptibility to external conditions (e.g., pressure, temperature). For a practical device it is desirable to minimize the influence of the external conditions on the performance. Thus, the single electrode configuration should be operated in an enclosed environment. 
Ultimately, a well-defined electrical response is expected either during operation under high vacuum conditions where the response is dominated by total electron yield or during operation in an enclosed environment with an ionizable medium, where the dominant contribution is due to the photoionization current (assuming moderate-to-low incident X-ray flux).

From results of the raster scans summarized in Fig.~\ref{fig:map} the electrical response is sensitive to geometry of the optical element
(signal enhancement at the edges) while the incident beam is not in direct contact with the electrode on the optical element. This observation suggests enhanced carrier generation at the edges and the presence of a charge transfer either in the surrounding medium or in the optical element.
The edge effect could be attributed to an increase in the total electron yield in grazing incidence to the edge surface which produces extra photoelectrons. These may drift in the surrounding medium towards the electrode if positive potential is applied and away from it if the negative potential is applied to the electrode. A drift of electric carriers can also occur in the optical element in the depletion region (where the electric field penetrates) if the carrier recombination lifetime is sufficient to traverse the distance between its origin of generation and the charge collecting electrode. Surface conductivity can also contribute to the final effect. 


The sensitivity of the raster scan to the sample geometry can be considered and utilized as an interesting variation of X-ray beam induced current (XBIC) imaging technique (e.g., \cite{Vyvenko02}), a variation which does not require metallization of the sample surface to be imaged.
The factors that form image contrast (besides X-ray absorption) may include the electric carrier mobility, distribution of the electric field and presence of inhomogeneities with enhanced edge surfaces. 


In summary, possibilities of forming an X-ray optical element with integrated non-invasive X-ray flux monitoring capability were considered. In particular, the function of such monitor in the photoemission mode was addressed by performing estimates of total electron yield from several X-ray optical grade materials. The obtained estimates for total electron yield were used to benchmark the experimental data collected under different experimental conditions. Electrical responses of single crystals of diamond and silicon were studied experimentally in a single electrode configuration with the electrode deposited on the entrance surface and not directly exposed by the incident beam. The experiments were conducted under ambient air at various levels of the incident X-ray flux. An additional experiment was carried out on an X-ray mirror in an enclosed environment. 
A non-zero contribution of photoemission to the electrical response was observed in the experiments with the single crystals using a high flux polychromatic hard X-rays.  
In contrast, the photoemission contribution to the response of the X-ray mirror can be neglected if the mirror is operated in an enclosed environment containing an ionizable medium and with moderate-to-low incident hard X-ray flux. Under these conditions photoionization of the medium can be utilized to form a flux monitoring optical element. It is concluded that practical implementations of X-ray flux monitoring optical elements in the single electrode configuration are possible in a high-vacuum or other enclosed and controlled environment. 

\section{Acknowledgements}
V.S. Bormashov, S.A. Tarelkin, Yu.V. Shvyd'ko, J. Terry, S. Ross and C. Jacobsen are acknowledged for helpful discussions on the topic. 
S.A. Terentyev and V.D. Blank are acknowledged for the diamond crystal provided.
Use of the Advanced Photon Source was supported by the U. S. Department of Energy, Office of Science, under Contract No. DE-AC02-06CH11357.  MRCAT operations are supported by the U.S. Department of Energy and the MRCAT member institutions.
\bibliographystyle{elsarticle-num}

\end{document}